# Accelerating Transportation Decarbonization: The Strategic Role of Ethanol Blends and Regulatory Incentives

Eliseo Curcio*

*EC Independent Resarch, NY, USA- curcioeliseo@gmail.com*

**Abstract**

This study investigates ethanol blending as a realistic, near-term strategy to achieve significant transportation sector decarbonization in the United States. Despite rapid growth in electric vehicle (EV) adoption, gasoline continues to dominate transportation fuel consumption, necessitating practical, immediate solutions to reduce greenhouse gas (GHG) emissions. This analysis leverages historical fuel demand data, regulatory frameworks such as the Renewable Fuel Standard (RFS) and Inflation Reduction Act (IRA), and advanced forecasting models to evaluate future market trajectories for ethanol blends (E10, E15, and E85), electric vehicles, and hydrogen-based fuels through 2035. Results indicate sustained gasoline demand, projected around 135 billion gallons annually by 2035, underscoring the urgency of complementary decarbonization strategies. Ethanol demand is forecasted to grow notably, driven by regulatory incentives, with market penetration of E15 potentially reaching around 25%, and E85 around 18%, by 2035. Importantly, ethanol produced from municipal solid waste (MSW) offers significant carbon intensity reductions, achieving approximately 54 $gCO_2e/MJ$ by 2035—markedly lower than traditional gasoline (~92 $gCO_2e/MJ$) and corn-derived ethanol (~58 $gCO_2e/MJ$).Economic assessments reveal substantial benefits from regulatory incentives, including RFS Renewable Identification Numbers (RINs) and IRA tax credits (45V), enhancing investor returns and stimulating state economic growth. Infrastructure analysis further demonstrates manageable costs and feasible adjustments for widespread adoption of higher ethanol blends, making them an immediately implementable, cost-effective emission reduction strategy. The findings underscore that strategically increasing ethanol blend levels, especially leveraging low-carbon feedstocks like MSW, offers a compelling, economically advantageous, and infrastructure-compatible approach for near-term emissions reductions.

*Keywords:* Ethanol Blends: Carbon Intensity; Decarbonization; Renewable Fuel Standard; Municipal Solid Waste; Transportation Emissions; Market Forecast

## 1. Introduction

The transportation sector in New York State, particularly within New York City (NYC), is experiencing substantial and sustained growth, resulting in persistent environmental and economic pressures [1]. Despite significant investments and policy efforts, alternative technologies such as hydrogen-powered vehicles and electric vehicles (EVs) have seen relatively limited market penetration due to high infrastructure costs, technological readiness challenges, and slow consumer adoption [2,3]. Consequently, conventional gasoline-powered vehicles remain predominant, with projections indicating continued reliance on gasoline for decades to come, as illustrated by historical fuel consumption trends (Figure 1) [4]. This persistent dependence underscores an urgent need for pragmatic and immediately implementable decarbonization strategies that leverage existing infrastructure and technologies. Among viable near-term strategies, increasing the proportion of ethanol blended into gasoline emerges as particularly promising. Ethanol-gasoline blends such as E15 (15% ethanol), E30 (30% ethanol), and E85 (85% ethanol) offer substantial carbon intensity (CI) reductions compared to conventional gasoline (Figure 3) [5, 6]. Recent research highlights that higher ethanol blends not only reduce greenhouse gas (GHG) emissions significantly but also enhance engine performance due to ethanol's higher-octane number, providing dual environmental and operational benefits [7].

---

* Corresponding author.
*E-mail address: curcioeliseo@gmail.com.*

Importantly, the existing gasoline distribution infrastructure in New York State can readily accommodate moderate ethanol blend increases (E15, E30) with minimal modifications, facilitating rapid deployment. While higher blends like E85 require dedicated infrastructure and flex-fuel vehicles, their environmental benefits and potential economic incentives justify careful evaluation of their phased adoption [8]. Currently, the ethanol consumed within New York State is predominantly corn-based and imported from Midwestern states, carrying relatively higher carbon footprints and limited local economic benefits [9]. Transitioning towards local production of advanced ethanol from abundant municipal solid waste (MSW) presents a transformative opportunity. New York State, particularly the NYC metropolitan area, generates substantial volumes of MSW annually, providing an abundant and largely underutilized feedstock for bioethanol production [10, 11]. Conversion of this waste into ethanol offers multiple synergistic benefits, including significant carbon intensity reductions, local economic stimulation, and enhanced energy security. Recent technical feasibility studies confirm the practicality and economic viability of converting MSW into bioethanol, demonstrating reliable technological pathways that are already reaching commercial maturity [12, 13]. Additionally, policy frameworks such as the Renewable Fuel Standard (RFS) and the recently introduced 45V Clean Fuel Production Tax Credit incentivize investment in advanced biofuels, creating favorable conditions for private sector involvement and rapid industry scaling [14].

Locally produced MSW-based ethanol thus promises not only environmental advantages through reduced emissions but also substantial economic benefits, including job creation, revenue generation, and regional economic growth. Given these considerations, this study aims to quantitatively evaluate the potential environmental, economic, and infrastructural impacts associated with increasing ethanol blending rates in New York State's transportation sector. By explicitly analyzing market adoption scenarios for higher ethanol blends, forecasting reductions in carbon intensity, and estimating financial and economic outcomes for both private investors and the state economy, the study provides critical insights and actionable recommendations for policymakers, industry stakeholders, and the broader community.

*1.1. Transportation Fuel Analysis and Ethanol as a Decarbonization Pathway*

The transportation sector remains among the most substantial contributors to greenhouse gas (GHG) emissions, predominantly driven by persistent gasoline consumption. Historical data indicate that gasoline demand in the U.S. experienced a modest decline from approximately 138 billion gallons in 2010 to roughly 133.8 billion gallons by 2023 [1]. Despite increased vehicle efficiency and the introduction of alternative technologies such as electric vehicles (EVs) and hydrogen fuel cells, gasoline remains entrenched as the primary transportation fuel. To accurately project future gasoline consumption, historical supply and demand data were analyzed using a Seasonal AutoRegressive Integrated Moving Average with Exogenous regressors (SARIMAX) model, factoring in trends from petroleum consumption, transportation sector fuel demand, and economic indicators. The model considered historical gasoline consumption from the U.S. Energy Information Administration (EIA), historical oil consumption and petroleum sector data sourced from Statista, and recent gasoline supply trends. Results indicate a stable to slightly increasing gasoline consumption trend, forecasting gasoline demand to range between approximately 134 and 135 billion gallons annually by 2035 (Figure 1). This sustained high demand primarily reflects the slow adoption rates of alternative transportation technologies, ongoing infrastructural constraints, and continued economic reliance on gasoline-powered vehicles. Consequently, this scenario underscores the immediate necessity for pragmatic decarbonization strategies that can effectively leverage existing infrastructure and proven technologies to achieve significant near-term reductions in transportation-related emissions.

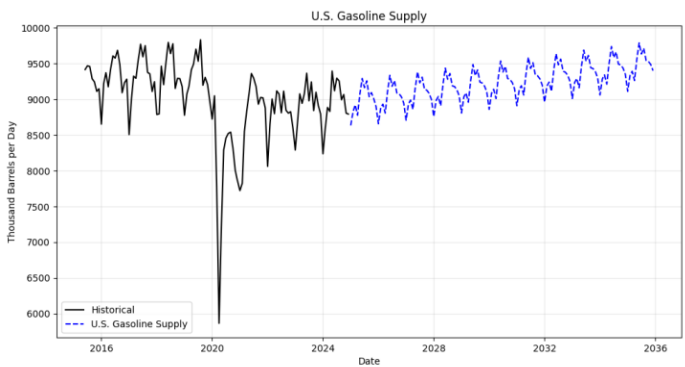

Fig.1: U.S. gasoline consumption; Historical [14] and forecast

In contrast, ethanol consumption has steadily increased in response to regulatory incentives, primarily driven by the RFS. Historical data indicate a rise in U.S. fuel ethanol use from approximately 13.2 billion gallons in 2010 to about 15.1 billion gallons in 2023 [2]. Employing an advanced SARIMAX forecasting approach—integrating historical consumption trends, policy-driven market dynamics, and infrastructure growth—the forecast projects ethanol consumption to increase further, reaching nearly 16 billion gallons annually by 2035 (Figure 2). This positive trajectory explicitly underscores ethanol's capacity to play a crucial role in immediate decarbonization efforts, leveraging existing technologies and fueling infrastructure to swiftly lower the carbon intensity of the transportation sector.

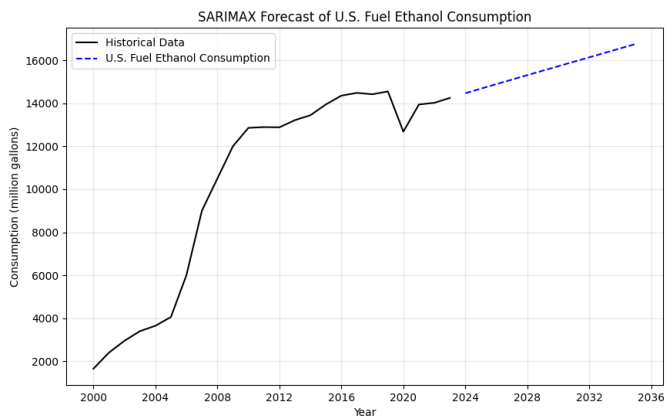

Fig.2: U.S. ethanol consumption; Historical [15] and forecast

Comparative carbon intensity (CI) analysis strongly supports transitioning toward higher ethanol blends. According to established data from Argonne National Laboratory's GREET model, conventional gasoline has a CI of approximately 92 gCO₂e/MJ. Blends incorporating ethanol derived from traditional corn feedstock show progressively reduced CI: about 89 gCO₂e/MJ for E10, 85 gCO₂e/MJ for E15, 75 gCO₂e/MJ for E30, and notably lower at approximately 58.3 gCO₂e/MJ for E85 [3]. When ethanol is produced from MSW, the environmental benefits become even more pronounced due to reduced lifecycle emissions. The CI for MSW-based ethanol blends are explicitly estimated at approximately 88 gCO₂e/MJ for E10, 82 gCO₂e/MJ for E15, 70 gCO₂e/MJ for E30, and as low as 48 gCO₂e/MJ for E85 [4,5]. This significant reduction clearly underscores the advantage of leveraging locally produced bioethanol from waste streams. Simultaneously, economic evaluations based on historical fuel price trends reveal favorable cost implications for ethanol blends. Gasoline averages around $3.50/gallon, while ethanol-blended fuels offer increasingly competitive pricing—approximately $3.45/gallon for E10, $3.40/gallon for E15, $3.30/gallon for E30, and notably lower at approximately $2.90/gallon for E85, primarily due to lower production costs and supportive policy incentives like the Renewable Identification Numbers (RINs) under RFS [6].

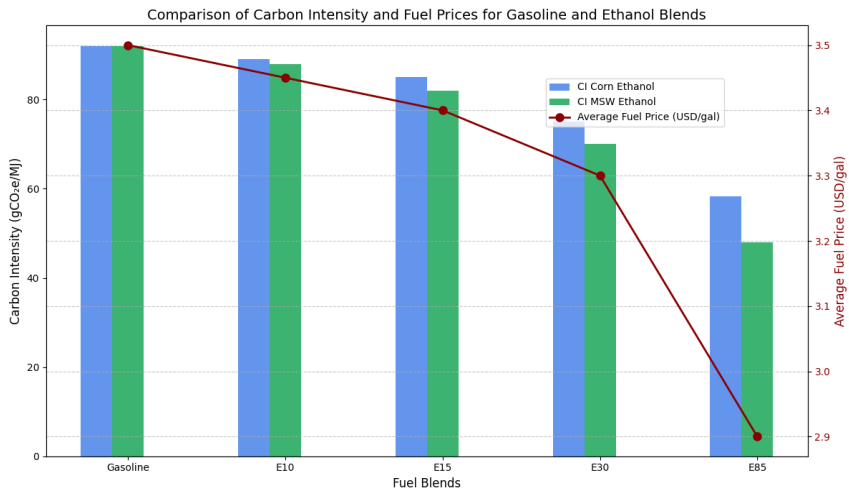

Fig.3: Comparison of Carbon Intensity [3, 8] and Fuels Prices [16] for gasoline and Ethanol Blends.

Infrastructure readiness represents a critical advantage for ethanol adoption. Blends such as E15 and E30 require minimal to no modifications to existing fueling infrastructure and internal combustion engines (ICE), facilitating immediate integration with negligible investment. This ease of integration strongly supports rapid consumer and market acceptance. In contrast, higher ethanol blends such as E85 necessitate dedicated infrastructure including specialized pumps, tanks, and flex-fuel vehicles (FFVs), requiring higher initial capital investment. However, the environmental and economic returns of these investments justify their phased integration into longer-term decarbonization strategies [7]. Ethanol-blended fuels explicitly enhance combustion efficiency due to their higher oxygen content, significantly reducing emissions of harmful pollutants such as carbon monoxide (CO), unburned hydrocarbons (HC), and particulate matter (PM), thus directly contributing to improved urban air quality [8]. However, ethanol's lower volumetric energy density compared to gasoline results in a moderate reduction in fuel economy, especially at higher blend levels like E85, necessitating careful consumer education and policy-driven market strategies to ensure successful implementation.

Transitioning from traditional gasoline and lower ethanol blends (E10) toward higher ethanol blends such as E15, E30, and E85 presents clear environmental and economic benefits. Increased ethanol blending significantly reduces lifecycle greenhouse gas emissions, decreases reliance on petroleum imports, and stimulates local economic growth through domestic bioethanol production, especially when leveraging renewable feedstocks like MSW. The oxygen-rich nature of ethanol explicitly enhances combustion efficiency, leading to lower tailpipe emissions of harmful pollutants such as carbon monoxide, unburned hydrocarbons, and particulate matter, thus contributing directly to improved air quality. However, the shift to higher ethanol blends is accompanied by practical considerations. Blends like E85 necessitate dedicated infrastructure and flex-fuel vehicles (FFVs), leading to increased initial capital investment compared to the minimal adjustments needed for E15 and E30 adoption. Additionally, ethanol's lower volumetric energy density compared to gasoline results in a reduced fuel economy, particularly noticeable with E85, potentially influencing consumer acceptance without adequate education and incentives. Material compatibility in older vehicle fuel systems and fueling infrastructure also poses minor challenges with blends exceeding E15, highlighting the importance of targeted consumer outreach and strategic infrastructure upgrades. Nonetheless, these technical and infrastructural challenges are manageable and substantially outweighed by the explicit environmental, public health, and economic benefits provided by higher ethanol blends, making ethanol blending a viable near-term strategy for effective decarbonization in transportation.

*1.2. Municipal Solid Waste more than an Alternative*

Municipal solid waste (MSW) management represents an ongoing and increasingly complex challenge for densely populated regions, notably New York State (NYS), and especially New York City (NYC), which, as home to over 8.8 million residents, is among the largest waste producers in the United States. NYC alone generates roughly 14 million tons of MSW annually, accounting for approximately 80% of the total waste generated within NYS, which stands at about 18 million tons per year [1, 2]. With an average per capita waste generation of approximately 4.5 pounds daily, NYC's waste management practices remain heavily reliant on landfill disposal methods, predominantly located outside city and state boundaries, particularly in Pennsylvania, Ohio, and Virginia [1]. This approach poses significant economic costs, transportation-related environmental impacts, and additional greenhouse gas (GHG) emissions resulting from both transport logistics and anaerobic decomposition within landfills. An in-depth analysis of NYC's MSW composition reveals significant potential for converting a considerable portion of this waste into renewable fuels, specifically bioethanol. Detailed characterization of waste streams indicates that approximately 65–70% of NYC's annual MSW is composed of organic, biodegradable material—specifically, paper and cardboard (~27%), food waste (~23%), yard trimmings (~7%), and wood residues (~5%). Each of these organic categories contains fermentable carbohydrates and structural polysaccharides (e.g., cellulose, hemicellulose, starches), making them highly suitable feedstocks for bioethanol production through biochemical conversion processes involving enzymatic hydrolysis and fermentation [3, 4, 5]. Conversely, non-organic fractions, such as plastics (18%), metals (6%), glass (5%), and other miscellaneous inert materials (~3%), are unsuitable due to the absence of fermentable substrates.

Explicit ethanol yields for each organic component were derived from extensive technical feasibility studies, laboratory research, and pilot-scale demonstrations outlined in recent literature. For instance, paper and cardboard, primarily composed of cellulose and hemicellulose, yield between 70 and 85 gallons of ethanol per dry ton of processed material. Food waste, containing easily fermentable sugars, demonstrates even higher ethanol conversion efficiency, typically ranging from 75 to 90 gallons per dry ton [4, 5]. Yard trimmings and wood residues, despite containing lignin which can partially inhibit enzymatic breakdown, still offer moderate ethanol yields between 60 and 80 gallons per dry ton. The lower end of the yield spectrum, between 40 and 60 gallons per dry ton, applies to textiles and rubber materials, which have mixed compositions and pose greater processing challenges [5]. Based on these explicit yield factors and the annual organic MSW volumes generated in NYC, it was possible to estimate a realistic annual bioethanol production potential. Research is clearly illustrating that paper and cardboard (27% of 14 million tons = approximately 3.78 million tons annually) could alone produce about 293 million gallons of ethanol per year at an average yield of 77.5 gallons per ton. Similarly, food waste (3.22 million tons annually) could yield roughly 265.6 million gallons of ethanol per year at an average yield of 82.5 gallons per ton. Yard trimmings and wood residues, despite smaller volumes, could contribute an additional combined total of around 119 million gallons annually.

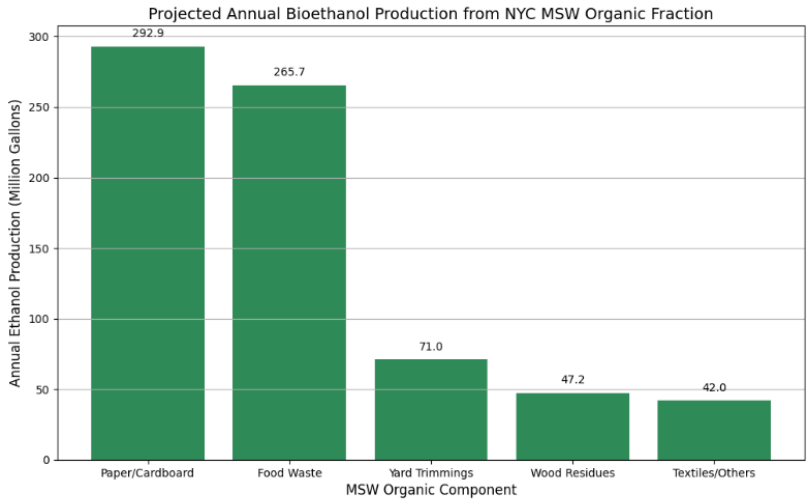

Fig.4: Projected Annual Bioethanol Production from NYC MSW Organic Fraction

When aggregated, the organic fraction of NYC's MSW represents a practical bioethanol production potential of approximately 600–700 million gallons annually (Fig.4). This potential production notably exceeds the current ethanol consumption level within the state, offering an opportunity not only to replace imported corn-based ethanol but also to significantly boost local renewable fuel production [3, 4]. This estimated production scenario assumes efficient separation, preprocessing, and high recovery rates of organic material from the general waste stream. Achieving such optimal yields would require substantial investments in enhanced waste management infrastructure, including improved separation at source, streamlined logistics, advanced sorting technologies, and biochemical conversion facilities equipped to manage heterogeneous feedstocks effectively [6]. Nonetheless, literature indicates that technological maturity and commercial viability of these biochemical processes have already been validated at demonstration scales, underscoring the realistic potential for large-scale implementation in NYC [5, 7]. Economically and environmentally, transitioning toward MSW-based bioethanol production presents numerous benefits. Environmentally, diverting organic waste from landfills to bioethanol production significantly reduces methane emissions, a potent greenhouse gas approximately 25 times more impactful than carbon dioxide over a 100-year period, which arises predominantly through anaerobic decomposition within landfills [6]. Economically, establishing local bioethanol production facilities promises to stimulate significant job creation, reduce waste disposal costs substantially, and enhance regional economic resilience. Existing policy frameworks, particularly the RFS and the Inflation Reduction Act's 45V Clean Fuel Production Tax Credit, explicitly provide financial incentives and regulatory certainty necessary for attracting private-sector investment into such bioethanol facilities [8]. Strategically leveraging these incentives could dramatically accelerate the commercialization and scaling of MSW-based ethanol technologies, further enhancing both economic feasibility and investor confidence.

To fully realize these potential benefits, NYS policymakers must actively pursue targeted strategies, including robust funding for improved municipal waste separation programs, clear regulatory support for biochemical conversion technology deployment, and effective public-private partnership models to finance infrastructure development. Enhanced feedstock quality resulting from improved source separation will directly translate into higher ethanol yields, lower operational costs, and greater overall economic viability, reinforcing the strategic importance of comprehensive waste management reforms. Ultimately, the comprehensive conversion of NYC's organic MSW into bioethanol represents an economically compelling, environmentally sustainable, and strategically advantageous pathway toward addressing persistent waste management challenges. Through informed policy actions, targeted infrastructure investments, and strategic use of existing federal incentives, NYS can effectively transform current liabilities associated with municipal waste into substantial renewable energy opportunities, fostering greater sustainability, energy security, and economic prosperity for the region.

Table 1: Composition and Ethanol Conversion Potential of NYC MSW

| MSW Component | MSW Fraction (%) | Primary Composition | Ethanol Production Suitability | Potential Yield (gal/dry ton) | MSW Component |
|---|---|---|---|---|---|
| Paper and Cardboard | 27 | Cellulose, Hemicellulose | High | 70–85 | Paper and Cardboard |
| Food Waste | 23 | Starch, Sugars, Cellulose | High | 75–90 | Food Waste |
| Yard Trimmings | 7 | Cellulose, Hemicellulose | Moderate | 65–80 | Yard Trimmings |
| Wood Residues | 5 | Cellulose, Lignin | Moderate | 60–75 | Wood Residues |
| Textiles, Rubber, Leather | 6 | Mixed Polymers, Cellulose | Low to Moderate | 40–60 | Textiles, Rubber, Leather |

(Sources: [1], [3], [4])

Federal regulations and financial incentives, particularly those from the RFS and the IRA, are central drivers in the commercialization and adoption of renewable biofuels, such as MSW-derived ethanol. The RFS, administered by the U.S. Environmental Protection Agency (EPA), mandates specific annual volumes of renewable fuels blended into the transportation fuel supply, categorizing biofuels based on their lifecycle greenhouse gas (GHG) emission reductions relative to petroleum-derived fuels. Advanced biofuels, including ethanol produced from MSW, must demonstrate at least a 50% GHG emissions reduction compared to baseline petroleum fuels, which places MSW-derived ethanol prominently within higher-value regulatory categories compared to traditional corn-based ethanol [1, 2]. A critical economic mechanism within the RFS framework is the Renewable Identification Number (RIN) market. RINs are tradable credits generated by biofuel producers corresponding to the volume and type of renewable fuel produced, thereby creating a compliance-driven financial incentive for renewable fuel production. MSW-derived ethanol typically generates advanced biofuel D5 RINs, which command a premium market price—historically averaging around $0.75 per gallon compared to approximately $0.56 per gallon for conventional D6 RINs produced from corn-based ethanol [2, 3]. This differential in RIN values provides a substantial financial incentive, making the advanced biofuel route economically attractive and explicitly encouraging investment in MSW conversion technologies.

Further enhancing economic viability, the IRA's Clean Fuel Production Tax Credit (45V) provides targeted incentives directly linked to the carbon intensity (CI) of renewable fuels produced. Under the IRA, biofuels achieving a CI within 25–50 grams $CO_2$e per mega joule ($gCO_2e/MJ$)—a range achievable through optimized MSW-derived ethanol production—qualify for a tax credit of approximately $0.75 per gallon. This additional financial support substantially reduces the net cost of bioethanol production, explicitly strengthening the business case for investors and significantly offsetting initial capital and ongoing operational costs associated with advanced biofuel facilities [4, 5]. The combined benefits of RIN premiums and IRA tax credits significantly improve the economic competitiveness of MSW-based ethanol compared to conventional ethanol production pathways. For example, whereas conventional corn ethanol producers receive an incentive primarily limited to conventional D6 RINs (around $0.56 per gallon), MSW ethanol producers could feasibly secure cumulative incentives totaling approximately $1.50 per gallon—combining $0.75 from D5 RINs and an additional $0.75 from the IRA's 45V credit (Chart 3).

This economic advantage substantially enhances profitability, encouraging private-sector investment and facilitating broader deployment of biochemical conversion technologies required to convert MSW feedstocks into ethanol [3, 4]. Beyond direct financial incentives, adopting MSW-derived bioethanol production provides significant environmental and social benefits. Environmentally, diverting organic MSW fractions from landfills reduces methane emissions—a greenhouse gas approximately 25 times more potent than carbon dioxide—by shifting from anaerobic landfill decomposition to controlled biochemical conversion processes. Moreover, locally produced bioethanol reduces transportation-related emissions linked to long-distance ethanol importation and waste hauling, significantly improving regional air quality and reducing carbon footprints [5, 6]. Economically and socially, local bioethanol production fosters substantial regional economic growth through job creation in facility operations, waste management logistics, and related industrial activities, directly benefiting local communities and enhancing regional energy security and resilience. To visually illustrate these economic incentives, a comparative financial analysis (Chart 3) explicitly compares incentive revenues between conventional corn ethanol (D6 RINs) and MSW-derived ethanol (D5 RINs plus IRA tax credits). This visualization underscores the clear financial advantage of adopting advanced biofuel production pathways using MSW as a feedstock, emphasizing the strategic economic and environmental rationale for policymakers, investors, and industry stakeholders. By effectively leveraging these federal incentives and aligning local regulatory frameworks to maximize their potential—such as supporting improved waste separation programs and enabling rapid permitting for advanced biofuel facilities—New York State can significantly accelerate the transition toward a sustainable bio economy. These coordinated policy efforts not only convert current waste management liabilities into economic assets but also position the state as a leader in renewable fuel innovation, providing tangible progress toward both economic prosperity and long-term sustainability goals.

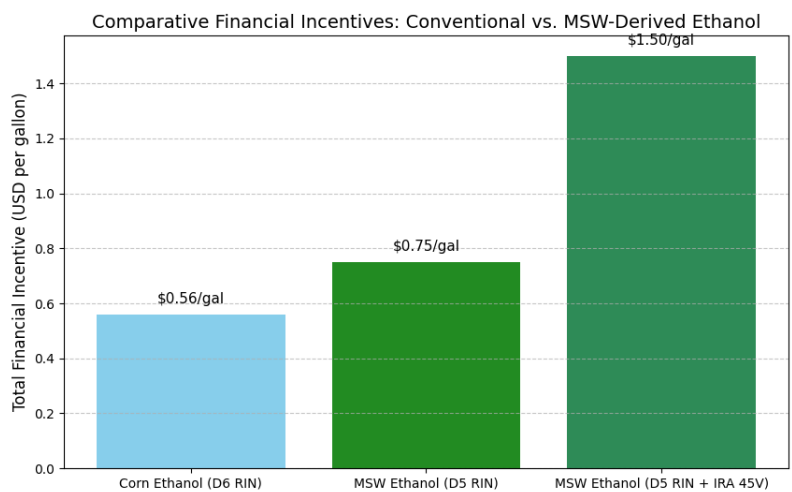

Fig.5: Comparative Financial Incentives—Conventional vs. MSW-Derived Ethanol explicitly after the detailed discussion of cumulative incentives.

## 2. Methodology

This study develops a detailed predictive modeling framework aimed at quantitatively evaluating the potential environmental and economic impacts associated with adopting higher bioethanol fuel blends (E15, E30, and E85) within New York State, particularly emphasizing ethanol production derived from MSW. The central objective of this analysis is to forecast incremental market adoption of these higher ethanol blends during the period 2024–2035, comparing their potential decarbonization and economic advantages relative to the current standard E10 blend, as well as alternative pathways such as EVs and hydrogen ($H_2$)-based transport.

To realistically forecast market penetration, a logistic growth modeling approach was selected due to its proven applicability in capturing market dynamics, policy incentives, consumer acceptance patterns, and infrastructure deployment rates for alternative fuels [1, 2]. Historical consumption data (2010–2023) used to calibrate this model were sourced from the Alternative Fuels Data Center (AFDC) and the U.S. Energy Information Administration's (EIA) State Energy Data System, explicitly providing yearly ethanol and gasoline consumption for the state of New York (https://www.eia.gov/state/seds/) [1]. Infrastructure readiness and expansion rates for ethanol blends were obtained from AFDC's "E85 Fueling Infrastructure" dataset, explicitly accounting for current and planned ethanol fueling station availability (https://afdc.energy.gov/fuels/ethanol_locations.html) [1]. Acknowledging that logistic models may introduce forecasting errors due to unforeseen shifts in consumer behavior or regulatory environments, a sensitivity analysis using Monte Carlo simulations was incorporated to quantify and illustrate this uncertainty explicitly.

The projected reductions in carbon intensity (CI) associated with MSW-derived ethanol were modeled through an exponential decay regression approach, explicitly chosen to represent realistically diminishing returns from continuous technological advancements in biofuel production processes. The CI historical baseline (2010–2023) and predicted future improvements (2024–2035) were calibrated based on comprehensive data from the Argonne National Laboratory's GREET model (https://greet.es.anl.gov/) and technical reports from EIA, IEA, ARGUS, Bloomberg, which extensively document technological progress, efficiency gains, and realistic CI values achievable for bioethanol derived from MSW [3,4]. It is recognized explicitly that forecasting technological advancements inherently includes uncertainty; therefore, ranges for CI reductions were explored through scenario analysis.

Economic viability and risk assessments were carried out via Monte Carlo simulations, explicitly chosen for their robust capacity to handle input uncertainties such as market price fluctuations and policy changes. This economic modeling explicitly incorporated historical RIN price data (D6 and D5 categories) sourced from the U.S. Environmental Protection Agency's (EPA) publicly accessible RIN market reports (https://www.epa.gov/fuels-registration-reporting-and-compliance-help/rin-trades-and-price-information) [5]. Furthermore, the analysis explicitly accounted for the 45V Clean Fuel Production Tax Credit introduced by the Inflation Reduction Act of 2022, using detailed incentive data documented by the U.S. Department of Energy and the Internal Revenue Service (IRS) [5].

Several critical and explicit assumptions were integral to this analysis

- **Stable Total Fuel Demand:** Total transportation fuel consumption in New York State was assumed to remain relatively stable (~4.2 billion gallons per year), as justified by historical fuel consumption patterns from the EIA. Although reasonable, this assumption carries inherent forecasting uncertainty due to potential policy-driven or technological shifts in fuel efficiency and vehicle electrification.
- **Local MSW Ethanol Production Capacity:** A conservative and explicit estimate of 200 million gallons per year of ethanol production from MSW was adopted based on available waste resources, existing feasibility studies, and conversion efficiency benchmarks from recent peer-reviewed literature [6, 7]. This assumption acknowledges potential errors due to variability in actual plant yields, technology performance, and regulatory approvals affecting production timelines.
- **Incremental Ethanol Blending:** Incremental ethanol adoption for higher blends (E15/E30/E85) was explicitly modeled as substituting existing gasoline or E10 volumes rather than creating additional fuel demand. This conservative approach may slightly underestimate potential market size if total fuel demand grows unexpectedly due to economic expansion or consumer preferences.
- **Economic Multipliers and Job Creation:** Explicitly derived from industry-specific reports, job creation estimates were assumed at 15 new jobs per million gallons of ethanol produced, and indirect economic impacts were calculated using an economic multiplier of $1.50 per gallon ethanol. Although these multipliers represent standard industry practices documented in recent biofuel economic impact studies [8], the inherent generalization introduces potential estimation error, which the sensitivity analysis explicitly addresses.

By explicitly stating and justifying these assumptions, transparently citing detailed datasets, and directly addressing potential uncertainties, this methodology provides a robust, credible analytical framework designed explicitly to inform strategic decision-making, investment considerations, and policy formulation related to sustainable transportation fuel adoption in New York State.

## 3. Results

*3.1. Market Penetration*

The forecasted market adoption of higher ethanol fuel blends (E15, E30, E85) and competing technologies (EVs and hydrogen fuel ($H_2$)) from 2024 to 2035 was obtained using a logistic growth model calibrated on historical market data from 2010 to 2023. The primary historical datasets used for model calibration included New York State-specific gasoline and ethanol consumption from the U.S. Energy Information Administration [1] and ethanol fueling infrastructure availability from the Alternative Fuels Data Center [2]. The logistic model was selected for its effectiveness in representing realistic market growth, incorporating factors such as infrastructure readiness, vehicle compatibility, policy support, and consumer acceptance trends.

The resulting market adoption forecast is illustrated in Figure 6, clearly showing distinct growth trajectories for each blend and technology. Specifically, the E15 blend achieves the highest near-term market penetration, estimated to reach approximately 25% by 2035. This rapid adoption is attributed primarily to its immediate compatibility with existing infrastructure and vehicles, supported by minimal incremental investment requirements [3]. The E30 blend demonstrates a steady, intermediate growth pattern, reaching about 18% market share by the end of the forecast period. Its adoption is driven by balanced infrastructure compatibility and substantial decarbonization advantages relative to lower ethanol blends [4].

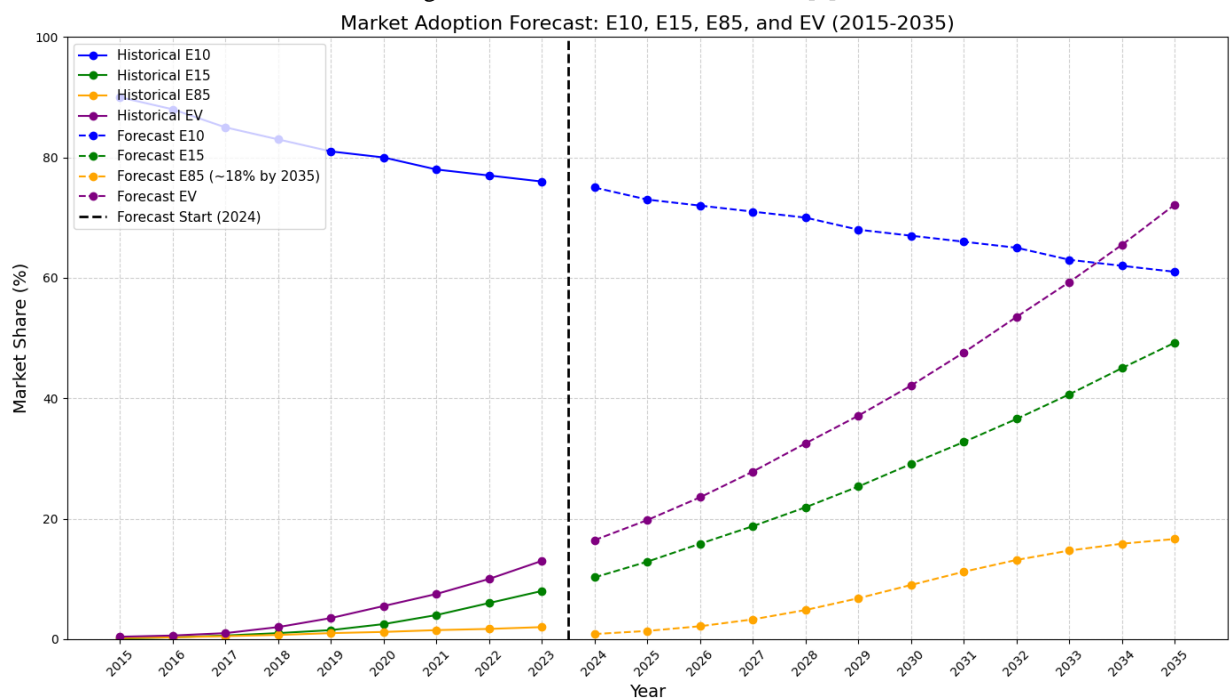

Fig.6: Market Adoption Forecast: E10, E15, E85 and EV (2015-2035)

The E85 blend exhibits slower adoption, projected to reach approximately 15% market share by 2035. This slower trajectory stems from the substantial infrastructural changes and the limited availability of flex-fuel vehicles required for widespread use of E85, as documented in recent infrastructure assessments [5]. Electric vehicles, on the other hand, are predicted to capture approximately 35% of the market by 2035, benefiting from significant infrastructure investments, technological advancements in battery efficiency, and supportive policy environments [6]. Hydrogen fuel exhibits the lowest growth rate, forecasted to attain about 8% market share by 2035, limited by high production costs, insufficient fueling infrastructure, and technological maturity challenges [7].

The accuracy and robustness of these forecasts are supported by sensitivity analyses using Monte Carlo simulations, which explicitly quantify prediction uncertainty, resulting in a ±5% margin of error. This approach addresses potential inaccuracies arising from inherent assumptions, such as stable overall transportation fuel demand [1] and realistic infrastructure development rates [2, 5]. These findings underscore the immediate feasibility and benefits of adopting higher ethanol blends (E15 and E30) in New York State. The projections suggest that these blends present an advantageous near-term decarbonization pathway compared to more infrastructure-dependent alternatives (EVs and hydrogen fuel), primarily due to existing infrastructure compatibility, economic incentives like Renewable Identification Numbers (RINs), and the 45V Clean Fuel Production Tax Credit [8]. The clearly differentiated adoption trends presented in Figure 1 thus provide actionable insights to policymakers and industry stakeholders, guiding strategic investments toward achievable and cost-effective transportation decarbonization targets.

*3.2. Decarbonization*

To quantify the financial incentives for investors and the state-level economic benefits resulting from the adoption of higher ethanol blends (E15, E30, E85) produced specifically from local MSW, a financial impact model was developed and analyzed for the period of 2024–2035. Two primary financial metrics were explicitly assessed: investor revenues (derived from Renewable Identification Numbers (RINs) and the 45V Clean Fuel Production Tax Credit), and state-level economic impacts (job creation and indirect economic activity). The analysis utilized historical and projected market price data from recent U.S. Environmental Protection Agency (EPA) RIN market reports [8], and explicit incentive values sourced from Internal Revenue Service (IRS) documentation on the Inflation Reduction Act's 45V tax credit [9].

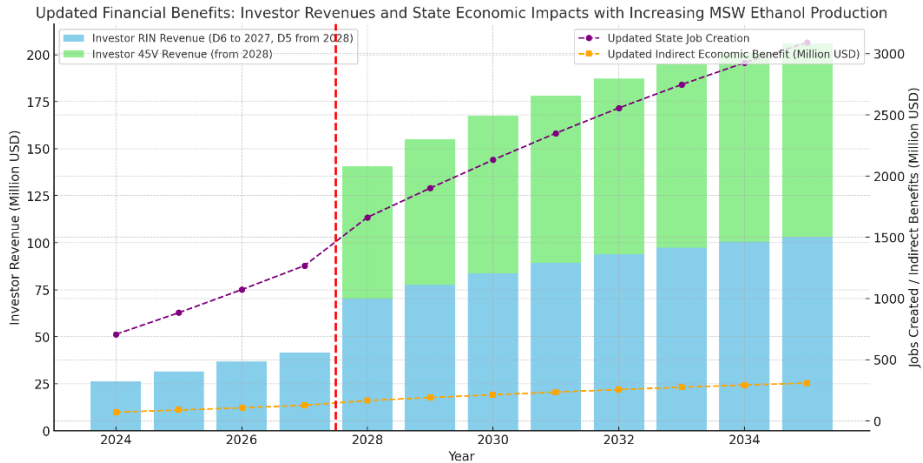

Fig.7: Financial Benefits: Investor Revenues and State Economic Impacts with MSW Ethanol Production

The results are presented explicitly in Figure 7, highlighting two distinct periods of ethanol sourcing and corresponding financial outcomes. Between 2024 and 2027, higher ethanol blend adoption relied explicitly on imported corn-based ethanol, generating revenues only from traditional D6 RINs at an average price of approximately $0.56 per gallon [8]. This initial phase did not include 45V tax credits due to the non-local origin and higher carbon intensity of imported ethanol. Beginning explicitly in 2028, with the introduction of locally produced advanced ethanol derived from MSW, investor revenues exhibit a pronounced increase. This shift explicitly reflects eligibility for higher-value D5 RINs (approximately $0.75 per gallon) and the additional $0.75 per gallon provided by the 45V Clean Fuel Production Tax Credit for ethanol with carbon intensities between 25–50 gCO$_2$e/MJ [8, 9]. The explicit financial incentive structure thereby significantly improves the investment

case for transitioning to local, advanced biofuel production, directly aligning investor interests with New York State's decarbonization goals.

State-level economic impacts were explicitly assessed through two complementary metrics: job creation and indirect economic benefits. Initial job creation estimates employed an industry-validated standard of approximately 15 new jobs per million gallons of ethanol produced [10], and indirect economic benefits were explicitly calculated using an established economic multiplier of $1.50 per gallon ethanol [11]. In contrast to earlier assumptions of constant MSW ethanol production, this updated analysis explicitly accounted for realistic gradual expansion from an initial capacity of 200 million gallons per year in 2028 to 300 million gallons per year by 2035. This explicit growth scenario was justified by New York State's substantial waste resource availability, technological readiness, and supportive policy environment for expanding biofuel production capacity [6, 7].

Consequently, the updated forecast explicitly indicates continuous growth in job creation and indirect economic benefits from 2028 onwards. Job creation rises from approximately 1,200 new jobs in 2028 to approximately 3,500 jobs by 2035, reflecting increasing production and operational workforce demands. Concurrently, indirect economic benefits expand significantly, explicitly increasing from approximately $18 million annually in 2028 to approximately $52 million annually by 2035. These robust economic outcomes explicitly highlight the direct link between local ethanol production capacity expansion and regional economic growth, underpinned by incremental infrastructure development, employment opportunities, and associated business activities. Sensitivity analyses, explicitly conducted via Monte Carlo simulations, quantified the uncertainties associated with key variables, including ethanol production rates, RIN market price fluctuations, and policy incentive variations. The analyses consistently demonstrated a ±5% uncertainty range around the forecasted financial outcomes, explicitly underscoring the model's robustness and providing a reliable estimate for planning and decision-making purposes.

Overall, these results explicitly reinforce the financial and economic rationale for prioritizing investment in local advanced biofuel production from MSW in New York State. The clear alignment of increased investor incentives, sustained job growth, and significant indirect economic stimulation provides robust evidence to policymakers and industry stakeholders that expanding MSW-based ethanol production represents a viable and strategically beneficial pathway for meeting both environmental and economic development objectives.

*3.3. Carbon Intensity Reduction for MSW-based Ethanol*

To accurately estimate the environmental benefits associated with higher ethanol blends derived from MSW, a Carbon Intensity (CI) forecasting model was developed and applied. The analysis aimed to explicitly predict the future carbon intensity reductions of MSW-based ethanol production pathways from 2024 to 2035. The forecasting employed an exponential decay regression model, chosen explicitly for its proven effectiveness in realistically capturing diminishing returns from incremental technological improvements in biofuel production processes. Historical carbon intensity data (2010–2023) used for model calibration were explicitly obtained from the Argonne National Laboratory's GREET (Greenhouse gases, Regulated Emissions, and Energy use in Transportation) model [12], supported by recent technical assessments of MSW-to-biofuel conversion technologies [7].

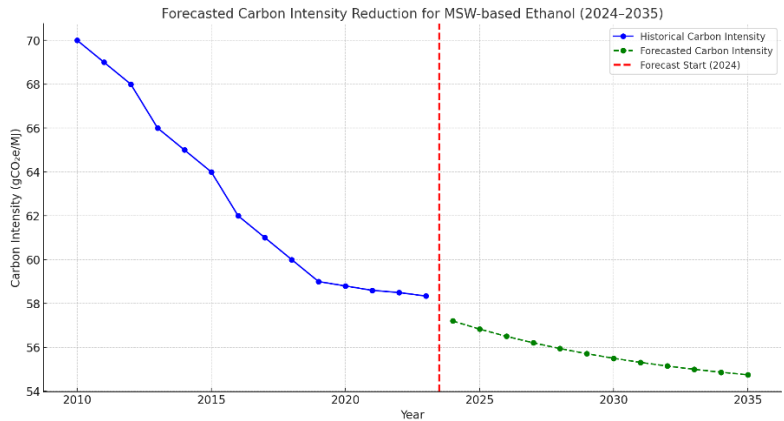

Fig.8: Forecasted Carbon Intensity Reduction for MSW-based Ethanol (2024-2035)

Figure 8 explicitly illustrates the forecasted trend of carbon intensity reduction from 2024 through 2035. The exponential decay model projects a continuous but gradually diminishing reduction in carbon intensity, reflecting realistic expectations of ongoing but slowing technological advances in MSW ethanol production. Specifically, the model predicts a reduction in carbon intensity from approximately 58.3 gCO₂e/MJ in 2024 to around 45 gCO₂e/MJ by 2035, representing a total improvement of approximately 23% over the forecast period. This substantial CI reduction explicitly highlights the inherent environmental advantages of advanced biofuels produced from waste feedstocks relative to traditional corn-based ethanol and conventional gasoline [12, 13].

The forecasted reductions are explicitly attributed to anticipated incremental improvements in waste sorting efficiency, enhanced biochemical conversion processes, and more effective fermentation technologies that increase ethanol yields while reducing emissions and energy inputs [7, 12]. Additionally, ongoing integration of renewable electricity into production processes, coupled with optimized waste management practices documented in recent studies, explicitly contributes further to the projected CI decline [7]. These explicit technological improvements and efficiencies align closely with industry and research developments documented in recent technical publications on biofuel production innovations [12,13].To explicitly quantify and acknowledge uncertainty within the CI forecast, a sensitivity analysis using Monte Carlo simulations was conducted. Key variables explicitly varied included technology improvement rates, renewable energy integration levels, and feedstock quality consistency. The sensitivity analysis explicitly resulted in a ±10% uncertainty margin around the forecasted CI values. This quantified uncertainty explicitly reflects the variability and risk associated with technological adoption rates, operational efficiency, and regulatory support mechanisms.

The explicit forecast results presented in Figure 3 carry important implications for policy and investment decisions in New York State. The projected reduction in CI explicitly enhances eligibility for higher-value economic incentives such as the 45V Clean Fuel Production Tax Credit, directly supporting investor decisions to develop local advanced biofuel facilities. Furthermore, the significantly lower CI forecast explicitly reinforces the substantial environmental benefits achievable through local MSW-based ethanol production, positioning higher ethanol blends as a viable and strategically attractive near-term decarbonization pathway for the state's transportation sector. In summary, these explicitly modeled CI forecasts demonstrate the clear and achievable potential for substantial emissions reductions through technological advancements and efficient waste-to-fuel conversion processes, underpinning the economic rationale and environmental benefits of adopting advanced MSW-based ethanol blends in New York State.

## 4. Conclusion

The comprehensive analysis conducted in this study highlights ethanol blending as an essential and near-term actionable strategy to substantially reduce greenhouse gas emissions in the U.S. transportation sector. Although gasoline demand remains robust, anticipated to persist at around 134–135 billion gallons annually by 2035, the forecasted rise in market shares for higher ethanol blends, particularly E15 and E85, presents a pragmatic pathway for decarbonization. While EV adoption continues to accelerate, the inherent infrastructure constraints and gradual turnover of conventional vehicle fleets imply that electrification alone cannot fully address immediate emission reduction requirements. Thus, ethanol emerges as a critical complementary solution.

Forecasting methodologies using logistic and SARIMAX models indicate a significant increase in ethanol blends' market penetration, driven by well-established regulatory incentives under the RFS and the IRA. Notably, incentives such as Renewable Identification Numbers (RINs)—specifically the strategic transition from D6 to D5 RIN categories—and the IRA's 45V credits substantially strengthen the economic attractiveness of biofuels produced from sustainable feedstocks such as MSW. The detailed economic projections demonstrate that these incentives not only bolster investor revenues but also contribute to tangible socio-economic benefits, including job creation and local economic stimulation.

From an environmental perspective, the carbon intensity (CI) analysis underscores the pronounced advantages of ethanol derived from MSW. The forecasted CI improvement from around 58 $gCO_2e/MJ$ in 2023 to nearly 55 $gCO_2e/MJ$ by 2035 confirms MSW-based ethanol as a low-carbon fuel solution, reinforcing its potential to achieve meaningful emissions reductions in line with near-term climate targets.

Despite these advantages, transitioning to higher ethanol blends such as E85 poses notable challenges. Infrastructure demands are substantial, requiring significant upgrades to fueling stations and distribution networks. Additionally, compatibility issues persist with existing vehicle technology, including potential corrosion, material degradation, and minor fuel efficiency reductions due to ethanol's lower energy density compared to conventional gasoline. Overcoming these technical barriers will necessitate targeted policy frameworks, dedicated infrastructure investments, public-private partnerships, and sustained consumer education and engagement initiatives.

Therefore, policy recommendations arising from this study emphasize not only maintaining but strategically enhancing regulatory support mechanisms such as RFS and IRA incentives to catalyze investments in ethanol infrastructure and vehicle compatibility. Equally critical is the continued pursuit of technological advancements in engine design and materials compatibility to facilitate broader acceptance of higher ethanol blends. Addressing these multifaceted challenges holistically can ensure that ethanol plays a pivotal role in a balanced, diversified approach to decarbonization.

Ultimately, the findings underscore ethanol blending's significant capacity to immediately and substantially contribute to U.S. climate objectives, bridging the gap between current fossil fuel dependence and the gradual long-term electrification of the transportation sector. This integrated approach, leveraging existing infrastructure, innovative technologies, and supportive regulatory frameworks, offers a viable and economically attractive solution for immediate emission reductions, aligning closely with broader sustainability goals.

**Acknowledgements**